\documentclass[referee]{aa}
\usepackage{graphics}
\newcommand{\Omo}{\Omega_{\rm o}}
\newcommand{\lamo}{\lambda_{\rm o}}
\newcommand{\Ho}{H_{\rm o}}
\newcommand{\Snu}{S_\nu}
\newcommand{\jnu}{j_\nu}
\newcommand{\To}{T_{\rm o}}
\newcommand{\Mgas}{M_{\rm gas}}
\newcommand{\fgas}{f_{\rm gas}}

\newcommand{\delc}{\delta_c}
\newcommand{\sigo}{\sigma_{\rm o}}
\newcommand{\Dg}{D_{\rm g}}

\newcommand{\Sobs}{S_\nu^{\rm obs}}
\newcommand{\fwhm}{\theta_{\rm fwhm}}
\newcommand{\fbeam}{f_{\rm beam}}
\newcommand{\msun}{\rm\,M_\odot}

\begin{document}

   \thesaurus{12.03.1;12.03.3;12.03.4;12.04.2;12.12.1;11.03.1} 
   \title{A Tale of Two SZ Sources}

   \subtitle{}

   \author{J.G.\,Bartlett\inst{1} \and A.\,Blanchard\inst{1} 
\and D.\,Barbosa\inst{1,2}}

   \offprints{J.G. Bartlett}
   \mail{bartlett@astro.u-strasbg.fr}
   \institute{Observatoire de Strasbourg,
	      Universit\'e Louis Pasteur,
	      11, rue de l'Universit\'e,
	      67000 Strasbourg,
	      FRANCE
	      Unit\'e associ\'ee au CNRS 
		({\tt http://astro.u-strasbg.fr/Obs.html})
	     \and Centro de Astrof\'{\i}sica da Universidade do Porto, 
	     Rua do Campo Alegre 823, 4150 Porto, PORTUGAL
             }

   \date{December, 23, 1997}

   \maketitle

   \begin{abstract}
	The recent discovery of two flux decrements in deep 
radio maps obtained by the VLA and the Ryle Telescope can
have powerful implications for the density parameter
of the Universe, $\Omo$.  
We outline these implications by modeling the decrements
as the thermal Sunyaev--Zel'dovich (SZ) effect from two clusters
{\em assuming} their properties are similar to those of 
the low redshift population.  In this case, the absence of any 
optical or X--ray counterparts argues that the clusters
must be at large redshifts.  We highlight the difficulty this poses 
for a critical cosmology by a comparison with a fiducial open 
model with $\Omo=0.2$ ($\lamo=0$).   
Applying the phenomenological X--ray luminosity--temperature 
relation needed to explain the EMSS cluster redshift distribution,
as inferred by Oukbir and Blanchard (1997), 
we convert the X--ray band upper limits to {\em lower} limits 
on the clusters' redshifts.  Comparison of the counts implied 
by these two SZ detections with model predictions, for clusters
with redshifts larger than these lower limits, illustrates
quantitatively the inability of the critical cosmology to account 
for such high--redshift clusters.  On the other hand, 
the open model with $\Omo=0.2$ remains consistent with the existence
of the two objects; it possibly has, however, difficulties with
current limits on spectral distortions and temperature fluctuations
of the cosmic microwave background.
The discussion demonstrates the value of 
SZ cluster searches for testing cosmological models and theories
of structure formation.
   \end{abstract}


\section{Introduction}

	Most favored theories of structure formation in the
Universe are based on the gravitational growth of initially 
small density perturbations with Gaussian statistics.  The
Gaussian characteristic finds its way into the mass function 
of cosmic structures and leads to the expectation of an
exponentially rapid decline in the number density of objects 
with increasing mass (Press \& Schechter \cite{PS}).  The 
interesting consequence is that the 
abundance of massive objects, such as galaxy clusters, is then
extremely sensitive to the power spectrum (amplitude and
shape) of the density fluctuations and  
their growth rate 
with redshift. Since the growth rate 
is controlled by the density parameter of the Universe, 
$\Omo$, this means 
that the {\em shape} of the redshift distribution of clusters of 
a given mass is sensitive to $\Omo$.  
In fact, once constrained by local data, the redshift distribution
depends {\em only} on the underlying cosmology, i.e., $\Omo$ (and to 
a lesser extent, on the cosmological constant $\Lambda$). 
In other words, the 
redshift distribution provides a probe of the density 
of the Universe (Oukbir \& Blanchard \cite{OB0}, \cite{OB}; 
Blanchard \& Bartlett \cite{b2}).
The sense is such that we expect many more clusters 
at large redshift if $\Omo<1$, because the 
growth rate is suppressed by the rapid expansion 
at late times in open models, leading to less evolution
towards the past. 

	Essential for the application of this probe of $\Omo$ 
is the existence of an easily observed cluster quantity  
well correlated with virial mass.  It has been cautioned
by many authors that the velocity dispersion of cluster member 
galaxies is too easily inflated by contamination of 
interlopers along the line--of--sight (e.g., Lucy \cite{interlop1};
Frenk et al. \cite{interlop2}; Bower et al. \cite{interlop3}).
Many have turned instead to the X--ray temperature
of the intracluster medium (ICM).  On theoretical 
grounds, it is believed that the gas is heated by infall
to the virial temperature of the cluster gravitational
potential well.  Numerical simulations in fact support 
the existence of a tight relation between virial mass and 
X-ray, which is to say, emission weighted, temperature
(Evrard \cite{Tsim1}; Evrard et al. \cite{Tsim2}).
The dependence of the relation is as expected 
based on the idea that the gas is shock heated to the
virial temperature on infall, although the simulations
indicate that there is an incomplete thermalization
of the gas, resulting in a temperature slightly 
($\sim 20\;$ \%) smaller than the virial value. 
The X--ray temperature is to be preferred
over the X--ray luminosity as an indicator of virial mass, because 
the X--ray luminosity depends not only on the 
temperature, but also on the quantity of gas and
on its density, or, what
is equivalent, the spatial distribution of the
ICM.  This spatial distribution is difficult to 
model, particularly because there is at present
no understanding of the origin of the ICM
core radius.

	Use of the X--ray temperature function to
constrain models of structure formation is rather
well developed as a subject (e.g., see Bartlett \cite{casa} 
and references therein).
Temperature data on clusters
at $z>0$ is just now becoming available, and
the possibility of even higher $z$ data from 
future space missions like XMM makes the application
of the redshift distribution test proposed
by Oukbir \& Blanchard a real possibility
over the near term (see, e.g., Sadat et al. \cite{SBO}).

	The Sunyaev--Zel'dovich (SZ) (Sunyaev \& Zel'dovich
\cite{SZ}) effect offers 
another, complimentary approach to the problem of 
applying mass function evolution as a probe of $\Omo$.  
Due to the distance independence of the 
surface brightness of the distortion, the effect
represents an efficient method of finding
high redshift clusters.  This should be
contrasted with X--ray emission, whose
surface brightness suffers the $(1+z)^{-4}$
cosmological dimming.  Moreover, as will be 
developed below, the SZ effect
has other, important advantages over X--ray studies:
the integrated SZ signal of a 
cluster, its flux density 
(measured in Jy), is proportional to the
{\em total hot gas mass} times the {\em particle
weighted} temperature.  This means that
the signal is independent of the 
gas' spatial distribution and that the
temperature involved is closely tied
to the cluster virial mass, by 
energy conservation during  
collapse, for it is simply the total
energy of the system divided by the 
number of gas particles.  For the
same reason, this temperature should also
be {\em much less sensitive to any temperature
structure in the gas} than the X--ray (emission weighted)
temperature. Thus, the SZ effect is an 
observable which combines ease of 
theoretical modeling with ease of detection at large $z$.  

	All of this has prompted several
calculations of the expected SZ number counts
and redshift distribution of SZ selected clusters,
and their dependence on the cosmological 
parameters and ICM evolution
(Korolyov et al. \cite{SZcounts1};
Bartlett \& Silk \cite{SZcounts2};
Markevitch et al. \cite{SZcounts3};
Barbosa et al. \cite{SZcounts4};
Eke et al. \cite{Tcon8};
Colafrancesco et al. \cite{SZcounts5}).  
The future of this kind
of study appears bright with the prospect
of the Planck Surveyor satellite mission 
({\tt http://astro.estec.esa.nl/Planck/}).
Ground based efforts have also made
astounding progress recently, and it
is already feasible to map $\sim 1$ 
square degree of sky to produce
number counts down to flux levels sufficient to test theories 
(Holzapfel, private communication).

	In this paper, we discuss what {\em may}
already be an indication of
clusters at very large redshift
and the resulting implications.
We refer to two radio decrements,
one found in a deep VLA field (Richards et al. \cite{vla:radio})
and the other detected by the Ryle Telescope
during an observation of a double
quasar system (Jones et al. \cite{ryle:radio}).  
Although these detections
await definitive confirmation, we will nevertheless 
proceed to outline here the
implications of their explanation as the thermal SZ effect
produced by two clusters.
What makes just two such objects 
of great importance is the
fact that no optical or X--ray 
counterparts have been observed, and
the flux limits in the X--ray are
so stringent that the clusters
would have to be at large redshift (Richards et al. 
\cite{vla:radio};
Kneissl \cite{kneissl1}; Kneissl et al. \cite{kneissl2}).  This is of 
paramount importance because, as we have mentioned, massive
clusters (say $M \geq 10^{15} $M$_{\odot}$) at large $z$ 
are not expected 
in critical models.  Our goal in this
paper is to quantify just how badly
critical models fare in this regard.
We emphasize that the modeling is based on
the observed characteristics of the galaxy
cluster population, in particular the X--ray
luminosity--temperature relation and constraints
on its potential evolution.  This excludes from
the present discussion the possibility of 
a large class of low luminosity clusters 
(both optical and X--ray).
We believe that a clear discussion in this 
restricted context is nevertheless useful.
(For this reason we dubb this work a ``tale''!).  
The procedure  
also demonstrates the great potential of SZ cluster
searches for constraining theories of structure
formation. 

	The plot of the tale proceeds as follows:  In the next
section, we introduce the two radio decrements and 
their properties which will be used later.  Then
we outline our modeling of the SZ 
cluster population and of the two 
radio decrements.  This is followed by 
a discussion of the X--ray emission to 
be expected from clusters producing the
observed SZ signals and the {\em minimum} 
redshifts imposed by the X--ray flux 
upper limits; this represents a key element
of our tale.  Finally, we discuss the results
and various caveats in the analysis before
bringing an end to the tale with a brief
summary.

\section{The Two Sources}

	One of the cluster candidates was discovered in a
deep VLA pointing of an HST Medium Deep Survey
field (Ri\-chards et al. \cite{vla:radio}).  Near the center of
the pointing, a radio flux decrement was detected
with an extension of around $30\times 60\;$ sq. arcsecs.  The 
integrated (negative) flux is about $-27 \pm 6.6\; \mu{\rm Jy}$
at the observation frequency of 8.44 GHz (recall that
Jy $= 10^{-23}$ ergs/s/cm$^2$/Hz).  
In what follows, we will scale all SZ fluxes to their
value at the emission peak of the effect, 
$\lambda=0.75$~mm.  In these terms, this object
is a source of $(4.2\pm 1)$ mJy.  The other possible
cluster was found during a Ryle Telescope (RT) observation
in the direction of the quasar pair PC1643+4631A,B 
(Jones et al. \cite{ryle:radio}), as
part of the telescope's ongoing effort to find
high-z clusters (Saunders \cite{ryle:prog}).  
This object is slightly larger
and stronger, corresponding to an integrated flux
decrement of $-410\pm 64\; \mu{\rm Jy}$ at 15 GHz and
extending over an area of $110\times 175\;$ sq. arcsecs.  Translating
to our fiducial frequency, we find a source of
$(19.64\pm 3)$ mJy.  

	The crucial aspect of these two candidates
is that, despite somewhat extensive efforts, no optical
or X--ray counterpart has been identified (Richards et al.
\cite{vla:radio}; Saunders et al. \cite{ryle:optical}).  
We will focus
on the implications of the X--ray data.  The VLA field
has been observed by the HRI aboard ROSAT, achieving
a limiting sensitivity of $\sim 2\times 10^{-14}\;$ ergs/s/cm$^2$
in the band [0.1,2.4]~keV.  A similar limit on the bolometric
flux (converted using $T=2.5~{\rm keV}$) has been set on 
the RT candidate (Kneissl \cite{kneissl1}; Kneissl et al. \cite{kneissl2}), 
in what was 
one of the last PSPC pointings
made with ROSAT.  Thus, in both cases, no X--ray emission
is detected down to very faint flux levels.

\section{Modeling the SZ Effect}

	In this section we discuss, first, our modeling of 
the cluster counts and redshift distribution as a function 
of the {\em total}, {\em integrated} SZ flux, i.e., assuming that
the clusters are unresolved.  This has the advantage
that the results are independent of the gas distribution,
as already remarked.  However, the actual observations
may in fact be resolving the supposed clusters, depending
upon their redshift.  We must therefore model the gas distribution
in order to interpret the observations.  We discuss
this aspect in the second subsection; we will argue 
that the singular isothermal sphere represents the 
most favorable case for a critical universe.

\subsection{SZ Counts and Redshift Distribution}

	Our notation and approach follow that
described in Barbosa et al. (\cite{SZcounts4}).  
The SZ surface brightness at position $\vec{\theta}$ is expressed as 
\begin{equation}
\label{SZbright}
i_\nu(\vec{\theta}) = y(\vec{\theta})j_\nu(x),
\end{equation}
where $x\equiv h_p\nu/k\To$ is a dimensionless frequency
expressed relative to the energy of the unperturbed CMB Planck
spectrum at $\To=2.728\;$ K (Fixsen et al. \cite{cmb:temp1}).  The 
function $j_\nu$ describes the spectral shape of the
effect:
\begin{eqnarray}
\label{jnu}
\nonumber
\jnu(x) = 2\frac{(k\To)^3}{(h_pc)^2} \frac{x^4\mbox{e}^x}{(\mbox{e}^x-1)^2}
	\left[\frac{x}{\tanh(x/2)} - 4\right]\\
\equiv 2\frac{(k\To)^3}{(h_pc)^2} f_\nu.
\end{eqnarray}
Planck's constant is written in these expressions as
$h_p$, the speed of light in vacuum as $c$, and
Boltzmann's constant as $k$.  Notice the introduction
of the dimensionless spectral function $f_\nu$.

	An integral of the pressure through the cluster (at 
position $\vec{\theta}$ relative to the center) 
determines the magnitude of the effect; this integral
is referred to as the Compton--$y$ parameter:
\begin{equation}
\label{comptony}
y \equiv \int \mbox{d}l \frac{kT}{m_ec^2} n_e \sigma_T,
\end{equation}
where $T$ is the temperature of the ICM (really, the electrons), 
$m_e$ is
the electron rest mass, $n_e$ the ICM electron density, 
and $\sigma_T$ is the Thompson cross section.  

	Finally, the quantity of primary interest to us 
is the total, integrated flux density from a cluster:
\begin{eqnarray}
\label{fluxdens}
\nonumber
\Snu(x,M,z) = \jnu(x) D_a^{-2}(z) \int \mbox{d}V \frac{kT(M,z)}{m_ec^2} 
	n_e(M,z) \sigma_T \\
\propto \Mgas <T>.
\end{eqnarray}
In this expression, $D_a(z)$ is the angular--size distance
in a Friedmann--Robertson--Walker metric -- 
\begin{eqnarray}
\label{Dang} 
\nonumber
D_a(z) = 2\Ho^{-1} c \left[\frac{\Omo z + (\Omo - 2)(\sqrt{1+\Omo z} - 1)}
	{\Omo^2 (1+z)^2}\right] \\
\; = 2\Ho^{-1} c D(z),
\end{eqnarray}
where we introduce the dimensionless quantity $D(z)$.
The Hubble constant is denoted by $\Ho$ and will also
be referred to by its dimensionless cousin 
$h\equiv \Ho/100$ km/s/Mpc.  Unless otherwise
specified, we use $h=1/2$.  
Notice that the integral over the virial volume
has reduced our expression to a product of the 
total gas mass times a temperature.  This is 
the origin of the statement that the effect
is independent of the ICM spatial distribution --
it depends only on the total gas mass.  The
formal definition of the temperature in the
expression is $<T>\propto (1/\Mgas) \int dV n_e T$,
which is the mean, {\em particle} weighted temperature,
i.e., the total thermal energy of the gas divided by the
total number of gas particles.  Even more so than
the X--ray temperature, which is emission weighted,
this quantity is expected to have a tight correlation
with cluster virial mass, just based on energy 
conservation during collapse; it is also much
less sensitive to any temperature structure in the ICM.
In addition, there is the well known fact that
the SZ surface brightness, Eq. (\ref{SZbright}),
is independent of redshift, assuming constant
cluster properties; thus, clusters may be
found at large redshift, 
whereas the X--ray surface brightness
suffers from `cosmological dimming'.
We emphasize the fact that SZ modeling
has these important advantages over modeling
based on cluster X--ray emission.

\begin{table}
\begin{center}
{\bf Table}\\
Model Parameters - normalized to the local X--ray temperature function$^a$ \\
\begin{tabular}{*4{|c}|}
\hline
$\Omo$ & $h$ & $\sigma_8$ & $n$   \\ 
\hline  
   0.2 & 0.5 & 1.37       & -1.10 \\
   1.0 & 0.5 & 0.61       & -1.85 \\
\hline
\end{tabular}\\
$a$ - Henry \& Arnaud (\cite{HA})
\end{center}
\end{table}

	Assuming the X--ray temperature and the SZ
temperature are the same, we may insert the T--M 
relation, tested by numerical simulations (Evrard \cite{Tsim1}; 
Evrard et al. \cite{Tsim2}), into
our flux density expression to obtain
\begin{eqnarray}
\label{SZfluxdens}
\nonumber
S_{\nu} = (8\,\mbox{\rm mJy}\, h^{8/3}) f_\nu(x) \fgas 
	\Omega_o^{1/3}M_{15}^{5/3}[\frac{\Delta(z)}{178}]^{1/3} \\
	(1+z) D^{-2}(z),
\end{eqnarray}
which depends on the total virial mass ($M_{15}\equiv M/10^{15}\;
\msun$),
the gas mass
fraction, $\fgas$, and the redshift of the cluster.  
Other quantities appearing in this equation are the 
mean density contrast for virialization, $\Delta(z,\Omo,\lamo)$
($=178$ for $\Omo=1$, $\lamo=0$), and the dimensionless functions
$f_\nu$ and $D(z)$ introduced in Eqs. (\ref{jnu}) and
(\ref{Dang}).  For the time
being, we will suppose that $\fgas=0.06\; h^{-1.5}$ 
(Evrard \cite {gasfrac}) and that it is constant
over mass and redshift; we will re--address this issue
later.  This simplifies the discussion
and will help to clearly distinguish the various important
physical effects.  

	We may now transform
the mass function, for which we shall adopt the 
Press--Schechter (1974) formula --
\begin{equation}
\label{psfunc}
n(M,z) dM = \sqrt{\frac{2}{\pi}} \frac{<\rho>}{M} \nu(M,z) 
	\left| \frac{\mbox{d}\ln \sigma(M)}{\mbox{d}\ln M} \right| 
\mbox{e}^{-\nu^2/2} \frac{dM}{M}
\end{equation}
-- into a number density of clusters at each redshift with 
a given SZ flux density.  In this way, we can calculate 
the integrated SZ source counts and the redshift distribution
of sources at a given SZ flux density.  In Eq. (\ref{psfunc}),
$n(M,z)$ gives, at redshift $z$, the {\em comoving } 
number density of collapsed objects with mass $M$, per interval
of $M$.  The quantity $<\rho>$ represents 
the {\em comoving} cosmic mass density
and $\nu(M,z)\equiv \delc(z)/\sigma(M,z)$, with
$\delc$ equal to the critical {\em linear} over--density
required for collapse and $\sigma(M,z)$ the 
amplitude of the density perturbations on a 
mass scale $M$ at redshift $z$.  More explicitly,
$\delc(z,\Omo,\lamo)$ and $\sigma(M,z)=\sigo(M)\times
(\Dg(z)/\Dg(0))$, $\Dg(z,\Omo,\lamo)$ being the linear growth
factor.  It is essentially through $\Dg$ that the
dependence on cosmology ($\Omo$, $\lamo$) enters the 
mass function (see, e.g., Bartlett \cite{casa} for a detailed
discussion).

\begin{figure}
\label{zdist}
\resizebox{\hsize}{!}{\includegraphics{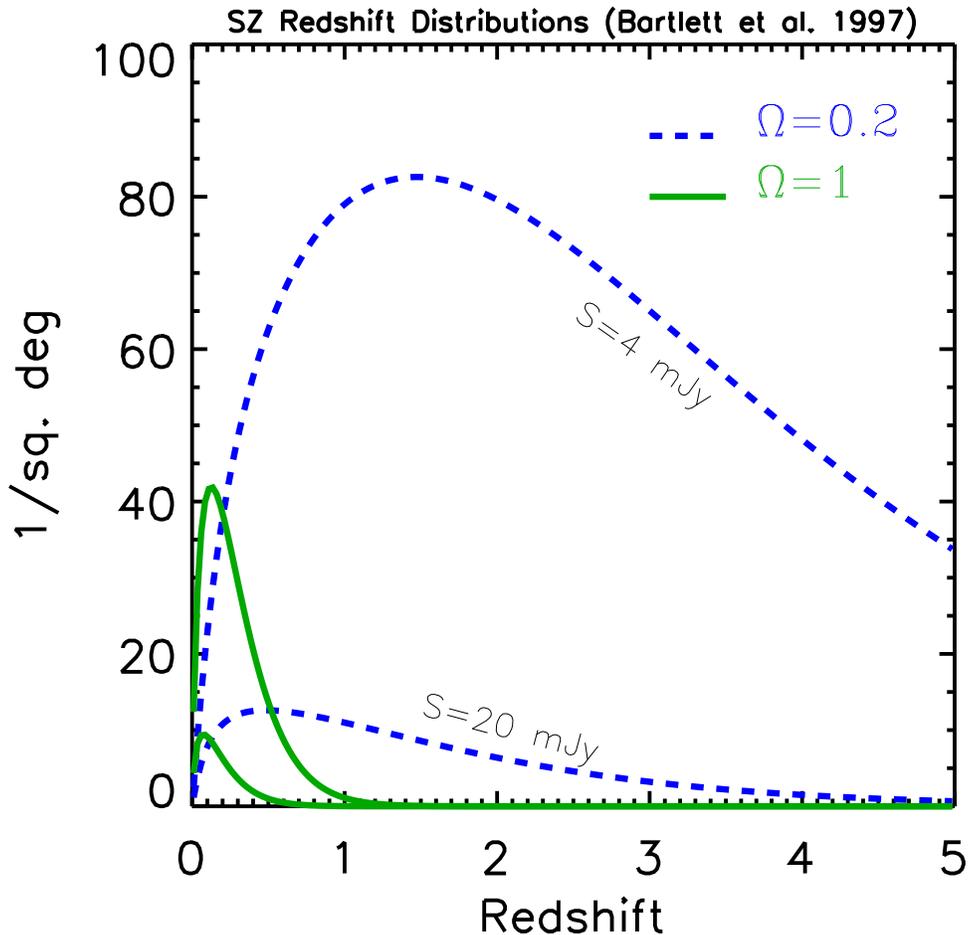}}
\caption{Redshift distribution of clusters with
flux densities of 4 and 20 mJy at $\lambda=0.75$~mm.
The dashed curves are for the open model.  Although 
for clarity not explicitly labeled,
the solid curves show the results for the critical model
for the same flux densities.  The model parameters are
given in the Table, and we use $h=1/2$.}
\end{figure}

	For illustration we will concentrate on the comparison
of a critical model ($h=1/2$) with an open model characterized
by $\Omo=0.2$ ($\lamo=0$ and $h=1/2$).  Both are normalized to the 
present--day cluster X--ray
temperature function.  The normalization
is performed by constraining, for each $\Omo$, the
amplitude, $\sigma_8$, and spectral index, $\alpha$, of 
an assumed power--law density perturbation power spectrum: 
$\sigo(M) = \sigma_8(M/M_8)^{-\alpha}$, where $M_8$ is the mass
enclosed in a sphere of radius $8h^{-1}$ Mpc.  In more standard
terms, $\alpha = (n+3)/6$, where $n$ is the spectral index
of the power spectrum: $P(k)\propto k^n$.  The result
of this normalization for the two models is given in the 
Table (Oukbir et al. \cite{OBB}; 
Oukbir \& Blanchard \cite{OB}).  

	The integrated cluster counts are shown 
in Figure 4 and discussed in Section
\ref{plot}, where we compare the results to the 
counts implied by the VLA and RT detections.  Here, in Figure 1,
we give the redshift distribution
of clusters of fixed flux density for the two
cosmologies.  For flux densities comparable
to those observed, we see a very large difference in
the predicted number of clusters at large redshifts.  
Thus, the existence of even
a small number of clusters at these
flux levels with redshifts beyond unity can lead 
to extremely strong constraints on $\Omo$.

\subsection{Observed SZ Signals}

\begin{figure}
\label{SZmassfig}
\resizebox{\hsize}{!}{\includegraphics{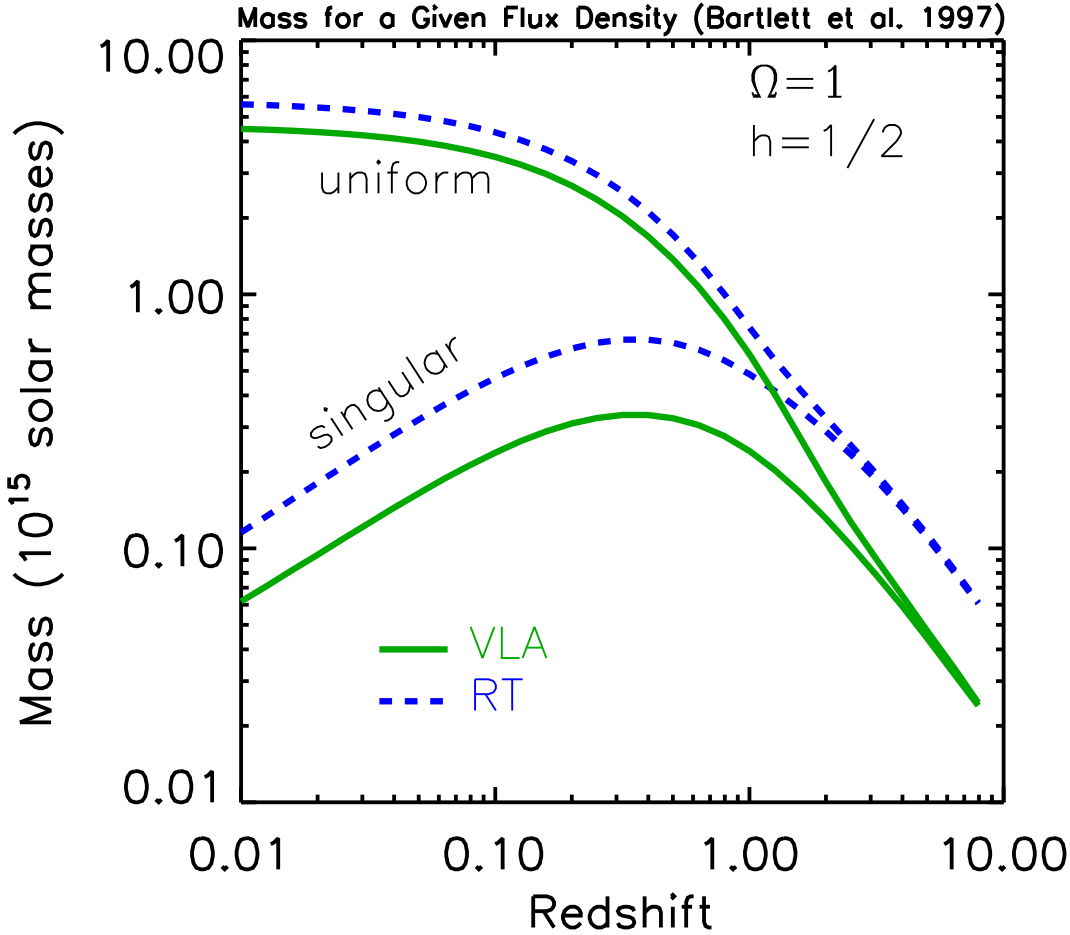}}
\caption{The virial mass required to explain each observed SZ source as a 
function of redshift; solid curves for the VLA, dashed for the RT.  
Gaussians beams of $\fwhm=1\;$ and
2 arcmins have been used for the VLA and RT, respectively. 
For each source, we show the results for $x_v=0\;$ and 500. The
latter approaches a singular isothermal sphere model.  In all cases
$h=1/2$ and $\Omo=1$.}
\end{figure}

	Now let us consider the problems associated with
the fact that the radio observations may be resolving
the cluster candidates.  We need to relate the 
observed, {\em resolved} SZ flux, $\Sobs$, to the {\em
total} flux integrated over the entire cluster volume, 
$\Snu$, which was the quantity considered in the 
previous subsection.  To do this, we adopt the 
isothermal $\beta$--model for the ICM density:
$n(r) \propto [1 + (r/r_c)^2]^{-3\beta/2}$.  The only
free parameter will be the core radius, $r_c$, because we will
fix $\beta=2/3$.  For the telescope beam profiles, 
we use Gaussians written 
in terms of the beam {\sf full width half maximum}, $\fwhm$:
$\fbeam(\theta) = \mbox{e}^{-4{\rm ln}(2) \theta^2/\fwhm^2}$,
where $\theta$ is the angle relative to the beam center.

	It is now possible to write 
down the observed fraction of the total SZ flux:
\begin{equation}
\xi \equiv \frac{\Sobs}{\Snu} = \frac
	{\int_0^{x_v}\mbox{d}x x (1+x^2)^{(1-3\beta)/2} \mbox{e}^{-4\ln(2)x^2
		(\theta_c(M,z)/\fwhm)^2}}
	{\int_0^{x_v}\mbox{d}x x (1+x^2)^{(1-3\beta)/2}}.
\end{equation}
We have introduced $x_v \equiv R_v/r_c$, where $R_v(M,z)$ is the
virial radius of a cluster of mass $M$ at redshift $z$.
Looking at $\Snu$ in Eq. (\ref{SZfluxdens}) 
and this expression for $\xi$, 
it is clear that $\Sobs$ is a 
function of $M$ and $z$.
Thus, as soon as the cosmological model and the characteristics of 
the instrument are  specified, we can find, for each redshift
and for given values of $x_v$ and $f_{\rm gas}$, 
the cluster mass required to produce 
the {\em observed} SZ flux.  The results for both 
observations, calculated using $\Omo=1$, are shown in 
Figure 2 for two extreme values of $x_v$. 
The telescope beams are described by $\fwhm=1$ arcmin and
$\fwhm=2$ arcmins for the VLA and RT, respectively;
in each case, the observed SZ source is taken to cover only one 
beam element.  As $x_v\rightarrow \infty$, i.e., as  $r_c\rightarrow 0$,
we recover the singular 
isothermal sphere profile, while $x_v\rightarrow 0$ describes an   
uniform density isothermal sphere, 
truncated at $R_v$.  Notice that because the 
singular profile concentrates more gas mass in the core, 
lower total masses are required to explain a given observation
($\Sobs$) than in the case of a uniform sphere.  The results
for $\Omo=0.2$ are similar, but the corresponding curves are 
roughly 2--3 times larger in mass.  What
goes into this calculation is simply the idea that the 
observed radio decrements are to be explained by hot gas  
heated to the virial temperature of collapsed objects.
Figure 2 indicates that, independent
of the SZ profile, the supposed objects
must have masses corresponding to groups or clusters of
galaxies, as was probably expected.

\begin{figure}
\label{Xflux_ev}
\resizebox{\hsize}{!}{\includegraphics{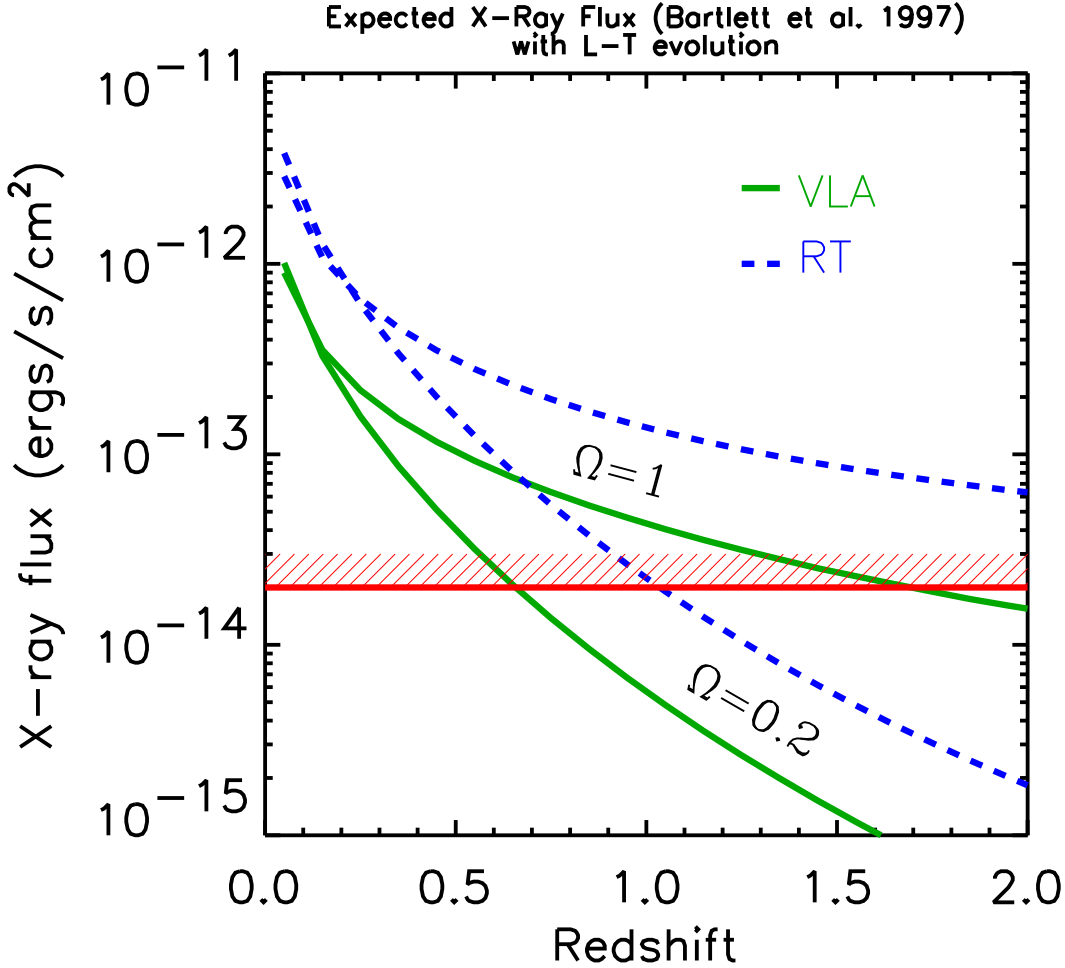}}
\caption{Expected X--ray flux for each SZ source 
using the evolution of $L(T)$ required to fit 
the EMSS redshift distribution (see text).
The X--ray limits on the two fields are numerically 
the same, at $2\times 10^{-14}\;
{\rm ergs/s/cm^2}$ (shown as the horizontal bar), 
but apply to the ROSAT [0.1--2.4]~keV band for
the VLA source (Richards et al. \cite{vla:radio}) and to 
the bolometric 
flux for the RT source (assuming $T\sim 2.5\;$ keV -- 
Kneissl et al. \cite{kneissl2}).
The solid curves for the VLA give the in--band flux, while the 
dashed curves for the RT correspond to expected bolometric flux.
As labeled, the upper pair of curves are for the critical
model and the lower pair are for the open model.}
\end{figure}

\begin{figure}
\label{Xflux_noev}
\resizebox{\hsize}{!}{\includegraphics{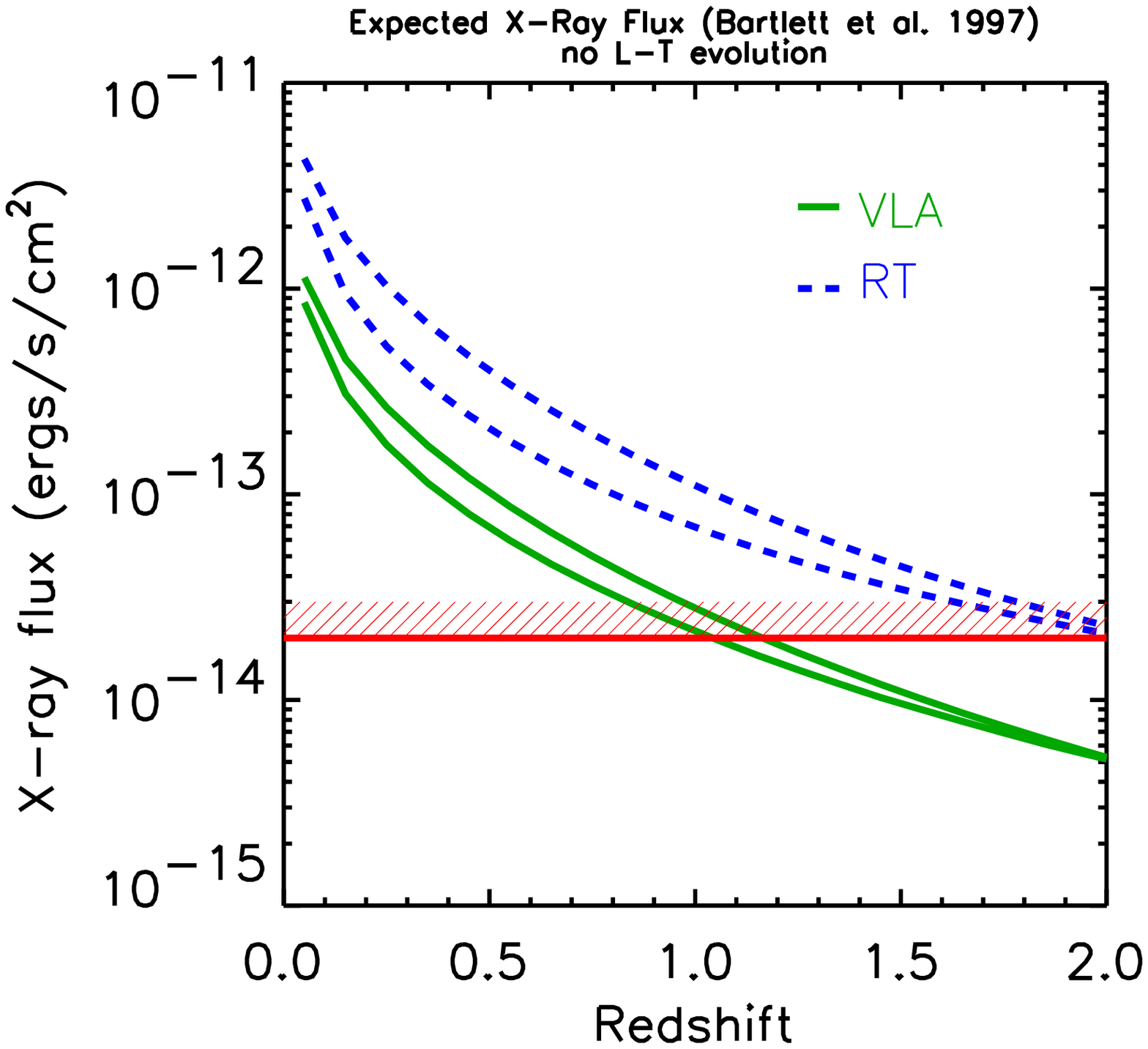}}
\caption{Same as Figure 3, but now assuming {\em
no evolution} of $L(T,z)$; the
observed $z=0$ relation is applied at all redshifts. 
In each case, the upper curve corresponds to the open 
cosmology, but there is very little difference between 
the two models.}
\end{figure}

\section{Modeling the Expected X--ray Emission - turning point of the
	plot}

	The limits on X--ray emission in the two SZ fields
are very stringent (and numerically similar): 
The ROSAT HRI places a limit of $f_x(0.1-2.4\; 
{\rm keV})< 2\times 10^{-14}\; {\rm ergs/s/cm^2}$
on the VLA field (Richards et al. \cite{vla:radio}),
and a new PSPC pointing puts a limit on the {\em bolometric}
flux on the RT field of $f_x < 2\times 10^{-14}\; 
{\rm ergs/s/cm^2}$ (assuming a $T\sim 2.5\; {\rm keV}$ -- 
Kneissl et al. \cite{kneissl2}).  These tight limits suggest that the 
(supposed) clusters are at large redshift.  Actually 
determining the minimum redshift thus imposed on each cluster
requires some additional modeling.  Fortunately,
there is a phenomenological approach:  as we have just
seen, the SZ flux tells us the cluster temperature (or 
mass) for any assumed redshift.  We wish
to associate an X--ray luminosity to this temperature
or, in other words, we are looking for a relation
$L(T,z)$.  Within a given cosmological model (i.e., 
given $\Omo$), this
relation is tantamount to specifying the redshift
distribution of a flux limited cluster catalog,
because it tells us the luminosity of clusters
of a given mass, whose abundance at any redshift
is given by the mass function. 
Oukbir \& Blanchard (\cite{OB}) have modeled 
the redshift distribution of the EMSS 
(Einstein Medium Sensitivity Survey, 
Gioia et al. \cite{EMSS})
cluster sample
and have shown that in fact the $L(T,z)$ relation can be used 
as a probe of $\Omo$ in this way, because once known,
$L(T,z)$ permits one to infer the evolution of 
the mass function (see introduction).
Although determined at $z=0$, the evolution of
$L(T,z)$ with $z$ is not yet fully constrained;
the potential to probe $\Omo$ using $L(T,z)$
is the motivation for current efforts to
constraint the relation at $z>0$ (Sadat et al. 1997).
For our purposes here, we will
use the $L(T,z)$ {\em required to match 
the observed redshift distribution of 
the EMSS} for arbitrary $\Omo$:
\begin{equation}
\label{LT}
L_x \propto (1+z)^{\beta}T^3
\end{equation}
with  
\begin{equation}
\label{LTbeta}
\beta \approx 4. \times \Omo - 3.
\end{equation}
(Oukbir \& Blanchard \cite{OB}; Sadat et al \cite{SBO}).

	We show the resulting X--ray flux 
for each SZ source in Figure 3.  For comparison, Figure
4 presents the results if we assume no evolution in the 
luminosity--temperature relation; in this latter
case, the X--ray flux is roughly the same for both
cosmological models.  Figure 3 shows that adding the
evolution required to fit the EMSS distribution --
Eqs. (\ref{LT}) and (\ref{LTbeta}) -- leads to a 
larger expected flux for the critical model.  This
is because clusters of a given temperature in the 
critical model must be {\em brighter} in the past 
to explain the observed number of clusters in the 
EMSS.  The opposite is true for the open model.  
In any case, and regardless of whether or not 
we incorporate evolution in the $L-T$ relation,
the X--ray limits imply that the cluster 
candidates must be at redshifts beyond unity
for the critical model; this is the limit
we will impose in the following.  We also remark
that this is consistent with the lack of any
optical cluster candidates in the two fields,
although a quantitative limit from the 
optical observations requires more detailed
consideration.  Looking at Figure 1, we expect the critical 
model to have difficulty accounting for the existence 
of clusters beyond a redshift of 1.  Massive clusters at 
high redshift are, on the contrary, expected to be relatively 
common in open models, and so the two decrements should
pose relatively little difficulty for our fiducial open model.  
We now quantify these comments.

\section{The Plot Comes Together}
\label{plot}

	We need to estimate the abundance of SZ 
sources of the kind observed.  
We will then be able to compare this abundance with
the predictions of the two models, and the important
element in this comparison will be the lower bound
on the redshift imposed by the X--ray (and optical) observations.
To start, we will assume that each telescope
simply observed blank regions of sky in a 
random fashion, making the statistics easier to handle.
One may argue that this applies to the VLA observations,
but not for the RT detection, which
deliberately pointed in the direction of a known
quasar pair.  Nevertheless, this will permit us to
quickly see the implications, and we can refine our
arguments later.  

	The VLA observations included two fields and found 
one SZ source.  Using a primary beam $\fwhm=312\;$ arcsecs, 
we find $\Omega^{\rm VLA}_{\rm obs}=0.018\;$ deg$^2$ for the 
total observed solid angle.  Assuming a similar 
sensitivity in both fields, this yields
an observed surface density of $\sim 50\;$ deg$^{-2}$ for SZ
sources with a flux density greater than 4.2 mJy at
our fiducial wavelength of $\lambda=0.75$ mm.  
Poisson statistics then tell us that the 95\%, 
{\em one--sided} confidence limits on the surface
density of these objects span the range 
$3 - 263\;$ deg$^{-2}$.  Following the same
line of reasoning for the RT, we note
that 3 fields where observed and one object was found;
a primary $\fwhm=6\;$ arcmins leads to 
$\Omega^{\rm Ryle}_{tot}=0.034\;$ deg$^2$ and a 
surface density range of $2 - 140\;$ deg$^{-2}$
(enclosed by the {\em one--sided}, 95\% confidence
limits) for sources with $\sim 20\;$ mJy at 
$\lambda=0.75\;$ mm.

\begin{figure}
\label{SZcounts}
\resizebox{\hsize}{!}{\includegraphics{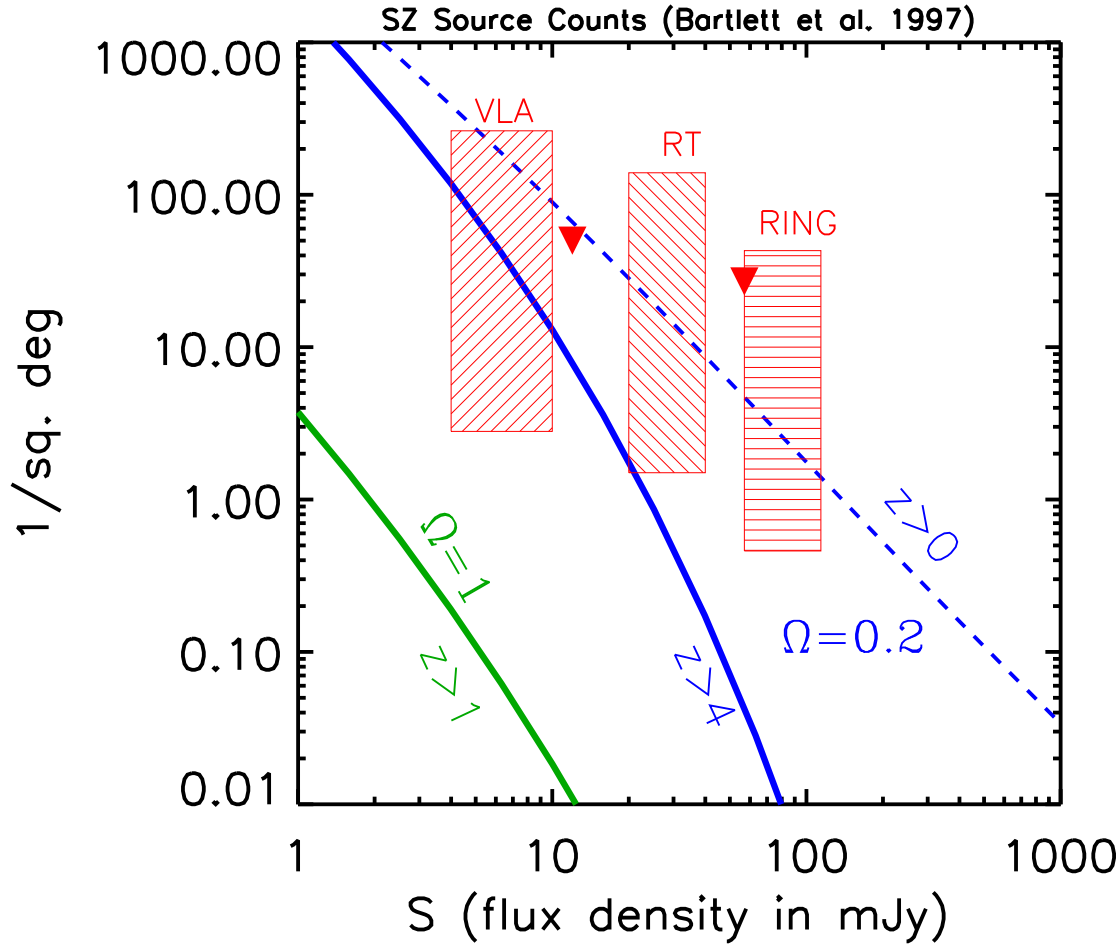}}
\caption{SZ source counts with observational constraints,
as a function of SZ flux density expressed at 
$\lambda=0.75\; {\rm mm}$.  The two hatched boxes on the left show the 
95\% {\em one--sided} confidence limits from the VLA and
the RT; due to the uncertain redshift of the clusters, 
there is a range of possible {\em total} SZ flux density,
which has for a minimum the value observed in each beam
and a maximum chosen here to correspond to $z=1$.  From 
the SuZIE blank fields, one can deduce the 95\% upper limit
shown as the triangle pointing downwards between the
VLA and RT boxes (Church et al. \cite{suzie}).  
The horizontally hatched 
box gives the constraints from the OVRO RING experiment
(Myers et al. \cite{RING}), assuming that there was
one cluster detected; the triangle at the far left of this
box shows the upper limit that results if, instead, one 
supposes that no clusters are present (see text).
We overlay the predictions of our fiducial 
open model ($\Omega=0.2$) for all clusters 
(dashed line) and for those clusters with $z>4$.  The critical model
has great difficulty explaining the observed objects
even with a lower redshift cutoff of only $z>1$; the actual
limit from the X--ray data could be stronger, but this would
fall well off to the lower left of the plot.  We assume
$h=1/2$.}
\end{figure}

	In Figure 5 we compare these
estimates with the predictions of the two models.
The critical model shown includes clusters
beyond $z=1$.  Even though this is rather generous 
for the critical model (looking at Figures 3 and 4), 
it fails by a large factor to explain
the counts indicated by the two radio decrements.
The open model counts are shown integrated 
upwards from $z=0$ and from $z=4$.   The open model can accommodate all the 
observational constraints (some of which are discussed 
below): clusters at $z \geq 1$ could easily produce these 
objects while escaping the X-ray limits.  The
critical model is incapable of explaining the 
observations, but the open model has little 
difficulty doing so.

\section{Discussion}

	The principal result of this work is given in Figure
\ref{SZcounts}, where we see clearly and quantitatively the 
difficulty faced by a critical cosmology if the VLA and RT radio
decrements are representative of the cluster population's SZ effect. 
We also show two other observational constraints in 
this figure.  The SuZIE instrument recently reported no 
detections down to 12 mJy (at $\lambda=0.75$ mm) on blank sky
covering $\Omega^{\rm SuZIE}_{tot}=0.06\;$ deg$^2$.
The resulting upper limit is shown in the figure as
the downward--pointing triangle between
the VLA and RT boxes.  The rightmost
box presents one possible interpretation of the results
of the OVRO RING experiment (Myers et al. \cite{RING}).
In this experiment, there is one field representing
a $>5\;\sigma$ fluctuation and for which no source has
yet been identified.  If we suppose that this fluctuation
is due to the SZ effect, then we deduce the constraints
given by the box based on this detection, at
$\sim 60$ mJy ($\lambda=0.75$ mm), over a total survey area
of $\Omega_{\rm obs}^{\rm RING}=0.1\;{\rm deg^2}$.  Once
again, we give the 95\% {\em one--sided} confidence limits,
and the range in flux density is from the detected flux 
density to the corrected, total flux density for $z=1$.
Alternatively, we could use the RING data as an upper limit
to the source counts, arguing that this one fluctuation
is not the result of a cluster.  We then obtain the
upper limit given as the downwards pointing arrow at the
far left-hand side of the RING box.

	In contrast to the critical model, our fiducial
open model has no difficulty in accounting for all of the
constraints shown in the figure.  It is the strong
upper limits on any X--ray flux from the objects, forcing
them to be at large redshift (Figure 3), that makes the distinction
between the two models not just a question of a factor of 
a few, but of {\em orders of magnitude}.  The
exponential behavior of the mass function provides
us with large ``leverage'' to discriminate models at
high $z$. To facilitate
the presentation in the figure, we have used a lower limit 
of $z>1$ on the critical model, but it should be emphasized
that the actual limit from Figures 3 and 4 is at least $z>2$
(for the RT),  
and this would put the curve completely off to the bottom 
left of the plot.
Thus, a straightforward interpretation of the results is that
the critical model is ruled out.

	Let us now discuss the various caveats to this 
`straightforward' interpretation.  The first thing that
must again be emphasized is that the line of argument relies
heavily on the idea that the radio decrements (are real) are
due to gas heated to the virial temperature of collapsed
objects and that these objects behave in a manner similar
to what is known about X--ray clusters.  It is perhaps
possible that the decrements are not due to such 
objects.  In this case, a different approach 
than the one presented here is needed.  The calculation
in this paper was made in order to understand what the radio 
decrements imply if one wishes to explain them by what 
we know as galaxy clusters.  

	Even within this context, 
there are several issues we should address.  
The first is the scatter in the $L(T,z)$ relation.
Arnaud and Evrard (1998) have shown that this scatter
is intrinsically smaller than previously thought, by
considering a cluster sample with high quality
X--ray observations.  The scatter is seen to be $~20\%$ 
around a relation $\propto T^{2.88}$ for non--cooling
flow clusters.  There are,
however, a couple of clusters below the mean relation
and well outside the scatter.  It is possible that
the SZ detections are the first indication of a larger
than expected population of intrinsically underluminous
clusters.

	Next, as already remarked, we chose for 
simplicity to hold the ICM gas mass
fraction, $\fgas$, constant over mass and redshift; one 
could imagine that it is in fact a function of both. 
Colafrancesco et al. (\cite{SZcounts5}) have investigated the SZ 
source counts including the possible effects of cluster 
evolution.  Accounting for cluster evolution in the present
context, however, will not substantially change the conclusion
in respect to the critical model -- we are already using, in
this model, a large gas mass fraction, as supported by
X--ray observations (Evrard \cite{gasfrac}) (This fraction
is in violation of primordial nucleosynthesis predictions for
$\Omo=1$, another problem for the critical model [White et al.
\cite{baryoncris}]).  Any reasonable
evolution would thus cause this fraction to decrease with either
mass or redshift, or both, thereby {\em decreasing} the 
counts predicted by the critical model -- things would get
worse.  One thing which could help the critical model is 
if the intergalactic medium surrounding the virialized
region of a cluster was heated to close to 1 keV.  One possible
mechanism could be the diffusion of electrons through the 
shock front that heats the gas to the virial temperature
(Chi\`eze et al. 1997). This
would increase the SZ flux density associated with a 
given cluster mass, pushing the predicted curves to the
right.  Such a mechanism could be in operation around
clusters, but the factor needed to reconcile the critical
model with the indicated counts seems unreasonable.  

	A particular source of concern is 
the fact that the RT object was found by observing known
quasar systems, and not a priori blank fields.  This
seems to be less of a worry for the VLA fields, because 
no double quasar systems were known prior to the observations 
(although one was subsequently found...).  We can get a feeling,
at least, for the effect of a possible bias on our results in 
the following manner: Supposing that instead of representing
random fields on the sky for a cluster search, double
quasar systems are {\em always} associated with clusters 
(which are responsible, say, for the two images).
Then another way to proceed with the RT
detection would be to take the observed sky density of 
such double quasar systems as the SZ source counts.
Over a survey area of $\sim 60\; {\rm deg}^2$, Schneider 
et al (\cite{quasdens}) found 90 quasars, 3 of which are
close double systems (with separations less than 
$\sim 400\;$~arcsecs).  One of these three is in fact
PC1643+4631, i.e., the RT field with the radio decrement.
The implied counts are then $\sim 0.05\; {\rm deg}^{-2}$,
with {\em one--sided 95\% confidence} upper and lower limits
of $\sim 0.13\; {\rm deg}^{-2}$ and $\sim 0.013\; {\rm deg}^{-2}$,
respectively.  Looking at Figure 5, we see that
this will not drastically alter the severity of the 
critical model's difficulty (remember that $z>2$ really
applies to the RT object). 

\section{Summary}

	In summary, we have seen the possible implications
of the recently discovered radio decrements in the VLA
and the Ryle Telescope.  If the spectra of these two objects
confirm an origin in the thermal SZ effect
from two galaxy clusters, then a critical model would
be in serious trouble.  A large
part of the importance of these two objects arises from 
the stringent X--ray and optical limits on the two fields.  This 
argues that the supposed clusters are at very large
redshift.  This has been the key,
because the SZ counts (the cluster abundance) at 
large redshift are enormously different between a critical
and an open model.  While such clusters are essentially non--existent
at $z>1$ in a critical model, they are to be expected in
open models, even at redshifts as large as 4.  

	However, all is not necessarily well with the open
model in light of constraints on spectral distortions
and temperature fluctuations of the cosmic microwave background.
Barbosa et al. (\cite{moriond}) show that an open model with
the power spectrum chosen here violates the FIRAS limit on $y$
(assuming a constant $\fgas$).  We may also estimate the 
{\em rms} temperature fluctuations created by the unresolved
cluster population by applying a P-D analysis (Condon \cite{con})
to our cluster counts.  We find that our fiducial open model 
here actually violates the present limit set by the SuZIE 
instrument (Church et al. \cite{suzie}). The significance of these 
shortcomings should be addressed with more 
careful modeling of the counts and predicted
fluctuations, and with a statistical comparison more appropriate 
to the non-Gaussian nature of the induced temperature fluctuations
(Bartlett et al., in preparation).

	Our modeling of the cluster population
is consistent and phenomenological in that it uses
the X--ray luminosity--temperature relation required
to explain the observed redshift distribution of 
EMSS clusters (Oukbir \& Blanchard \cite{OB}; Sadat et al.
\cite{SBO}) to find the expected X--ray flux and deduce
the corresponding redshift limits.  The discussion has
focused on the comparison between a critical model and a fiducial 
open model.  Such a simple comparison just to highlight the possible
implications seems justified at this juncture due to the
preliminary nature of the data, and the procedure demonstrates
the value of cluster SZ searches.  We await future results
with anticipation, noting for now that it is already 
possible to perform an SZ survey of $\sim 1\;$ square 
degree (Holzapfel, private communication; Holzapfel et al.
\cite{suzieinst}).

\begin{acknowledgements}
      
\end{acknowledgements}
We would like to thank G. Djorgovski for a discussion concerning the
abundance of close quasar pairs.  D.B is supported by the Praxis XXI CIENCIA-
BD/2790/93 grant attributed by JNICT, Portugal.

\end{document}